\documentclass{eas}
\usepackage{epsfig}
\def\dML{\mbox{$\nabla_{\ell} \Upsilon$}}

\begin{document}

\TitreGlobal{Mass Profiles and Shapes of Cosmological Structures}

%%-----------------------------
%%      the top matter
%%-----------------------------
\title{Mass-to-light ratios of ellipticals in $\Lambda$CDM}
\author{Napolitano N.R.}\address{INAF-Observatory of Capodimonte, Naples, Italy}
\author{PN.S team}\address{Planetary Nebula Spectrograph (PN.S) team: see {\it http://www.astro.rug.nl/$\sim$pns/pns\_team.html\#coll}}
\runningtitle{M/L of ellipticals in $\Lambda$CDM}
\setcounter{page}{23}
% Keep this line, even if the page will be settled afterwards..
\index{Author1, Napolitano N.R.}
\index{Author2, PN.S team}
%\index{Author3, C.}
% Repeat the authors here, this will help to make the final index

%
\begin{abstract}
We use the mass-to-light gradients in early-type galaxies to infer the global dark matter fraction, $f_{\rm d}=$M$_{\rm d}/$M$_{\rm *}$, for these systems. We discuss implications about the total star formation efficiency in dark-matter halos and show that the trend of $f_{\rm d}$ with mass produces virial mass-to-light ratios which are consistent with semi-analitical models. Preliminary kurtosis analysis of the quasi-constant M/L galaxies in Romanowsky et al. seems at odd with Dekel et al. simulations. 
\end{abstract}

\maketitle

%
%%-----------------------------
%%      your text
%%-----------------------------
\section{Introduction}

Recent finding of a sample of galaxies with a poor dark matter content within several effective radii (Romanowsky et al. 2003, R03 hereafter) seems to challenge the current paradigm of galaxy formation in  $\Lambda$CDM  cosmology where galaxies are seen to evolve in extended halos of dark matter with cuspy density profile (Navarro et al. 1997). Systems with a significant dark-matter component are expected to have mass-to-light ratios (M/L or $\Upsilon$ hereafter) deviating from the values typical of their stellar population. On the contrary R03 have shown that the kinematics of three intermediate luminosity galaxies can be modeled with a quasi--constant M/L consistent with the one expected from stellar population synthesis models.

This unexpected result have produced different interpretation either in the $\Lambda$CDM framework (Dekel et al. 2005 $=$ D05) or in MOND theory (Milgrom \& Sanders 2003). In particular, D05 address very radial stellar orbits and projection effects in order to explain declining velocity dispersion profiles.  

On the other hand, \cite{nap05} (N05 hereafter) made predictions of gradients of mass-to-light ratios (\dML\ hereafter) in early-type galaxies and have ascertained that ``quasi-constant M/L'' are indeed expected within $\Lambda$CDM, albeit the R03 sample shows \dML\ which are too low and conflicting with acceptable star formation efficiency and baryon fraction.
Low M/L gradients systems are a part of a generalised trend of \dML\ with stellar mass/luminosity
%: brightest galaxies have a larger fraction of dark-
%to-luminous matter, $f_{\rm d}=$M$_{\rm dm}/$M$_{\rm *}$ where M$_{\rm dm}$ and M$_{\rm *}$ are the dark and the luminous matter respectively, with respect fainter galaxies. 
%There are hints that this trend is reversed toward very faint (dwarf) galaxies which are thought to be dark-matter dominated.\\ 
%In \cite{nap05} we have interpreted such a trend as 
%which is the reflection of the global star formation efficiency being a function of the total stellar mass 
(see Section 3).\\
Here we show some preliminary results on the PNe's anisotropy of the R03 sample and compare them with the models from D05 and finally discuss whether the trend of the star formation efficiency is consistent with virial M/Ls.  
\begin{figure}[t]
   \centering
  \vspace{-0.4cm}
\hspace{-0.2cm} \epsfig{file=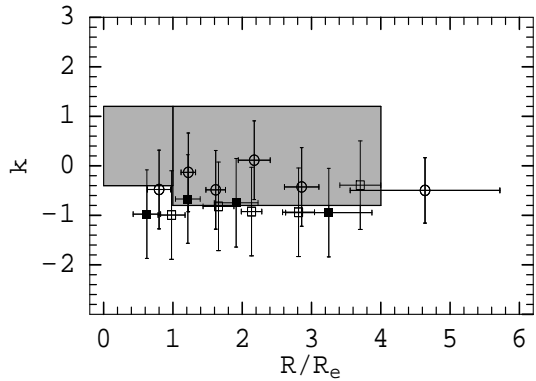,width=7cm,height=5.cm,angle=-0}
 \caption{Reduced kurtosis of NGC821 (full squares), NGC3379 (open squares), and NGC4494 (circles) compared to the average values from Dekel et al. 2005 (shaded boxes).}
       \label{fig1}
   \end{figure}
\section{Kurtosis profiles of PNe and anisotropy}
Simulations from D05 of galaxy merging of equal mass progenitors have shown that radial anisotropy naturally raises in the outskirts of merging remnants, where positive $h_4$ parameters should be measured. By converting $h_4$ to kurtosis using the approximated formula as in \cite{vdMF93}, it is possible to compare PN kurtosis to this predictions. 
In Fig. \ref{fig1} we show the radial kurtosis profiles of galaxies NGC821, NGC3379 and NGC4494, obtained with the latest PN.S data reduction run compared with the quoted average values from Dekel et al. (represented as shaded boxes). It looks like the observed values, albeit consistent within the errorbars, are systematically lower than the prediction from D05, which is already an indication of non strong radial anisotropy {\it di per se}.
\begin{figure}[t]
   \centering
  \vspace{-0.3cm}
\epsfig{file=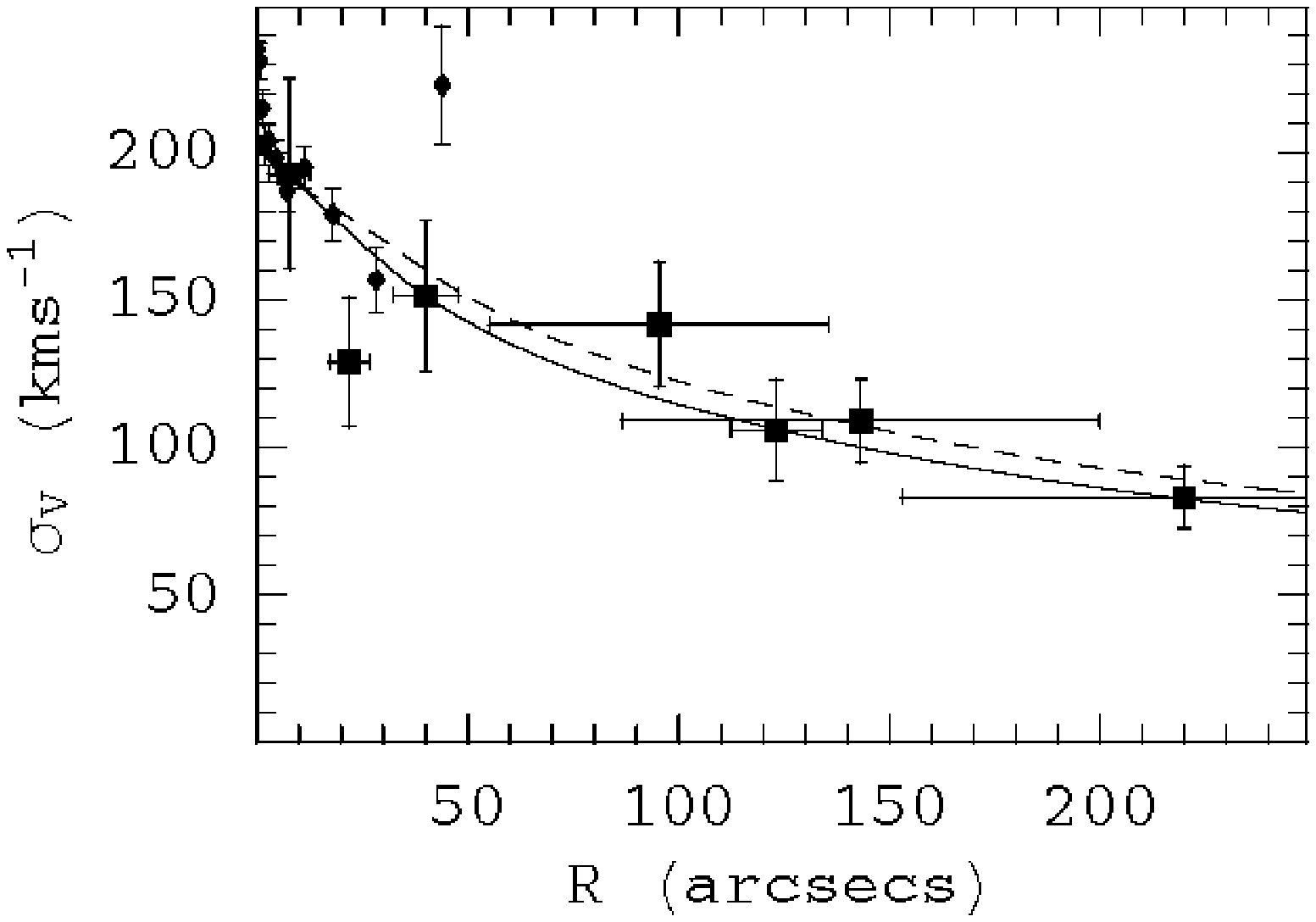,width=5.5cm,height=4.2cm,angle=-0}
\hspace{-0.1cm} 
\epsfig{file=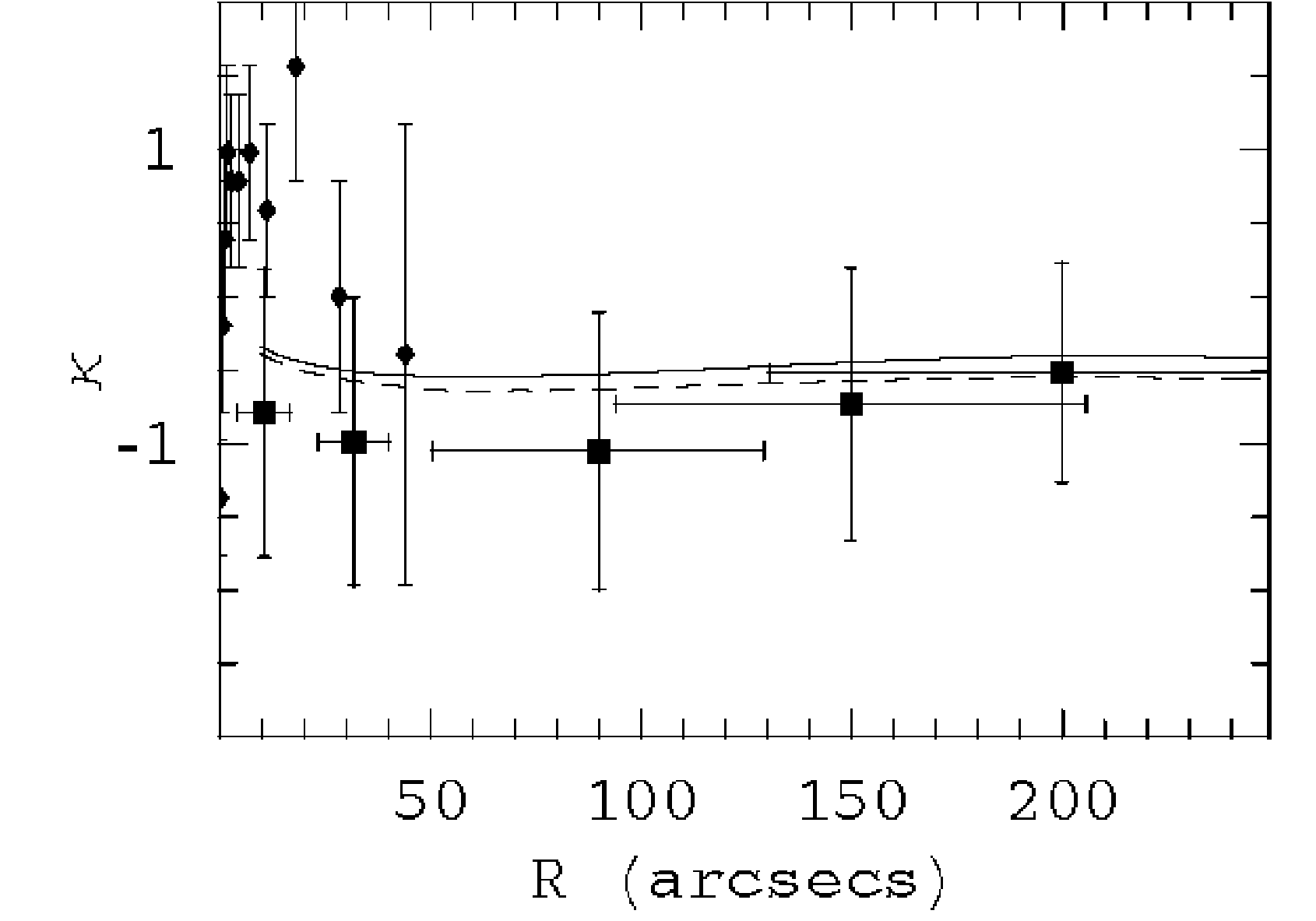,width=5.5cm,height=4.2cm,angle=-0}
 \caption{Left: Velocity dispersion of NGC821 from long-slit (diamonds) and PNe (full squares) with best fit model without dark halo assuming isotropy (solid line) and $\beta=-0.33$ (dashed line). Right: reduced kurtosis of NGC821, symbols as center panel.}
       \label{fig2}
   \end{figure}
A direct estimate of orbital anisotropy needs data modeling: following \cite{loma03} we have constrained the anisotropy parameter with Jeans analysis using 2nd and 4th velocity moments. Fig. \ref{fig2} shows the best-fit to the velocity dispersion and kurtosis for the case without dark-matter: best-fit parameters are  $\Upsilon_{*,B}=12.8\pm1.0$ and $\beta=-0.33\pm0.12$. Radial anisotropy ($\beta>0$) is marginally allowed when forcing the models to a stellar mass-to-light ratio $\Upsilon_{*,B}=10$ in order to accommodate a dark halo. For this case halo concentration is $c=5$, i.e. much lower then expected from N-body simulations (Bullock et al. 2001).
\section{Global dark-matter fraction and star formation efficiency}
\begin{figure}[t]
   \centering
  \vspace{-0.5cm}
\hspace{-0.25cm} 
\epsfig{file=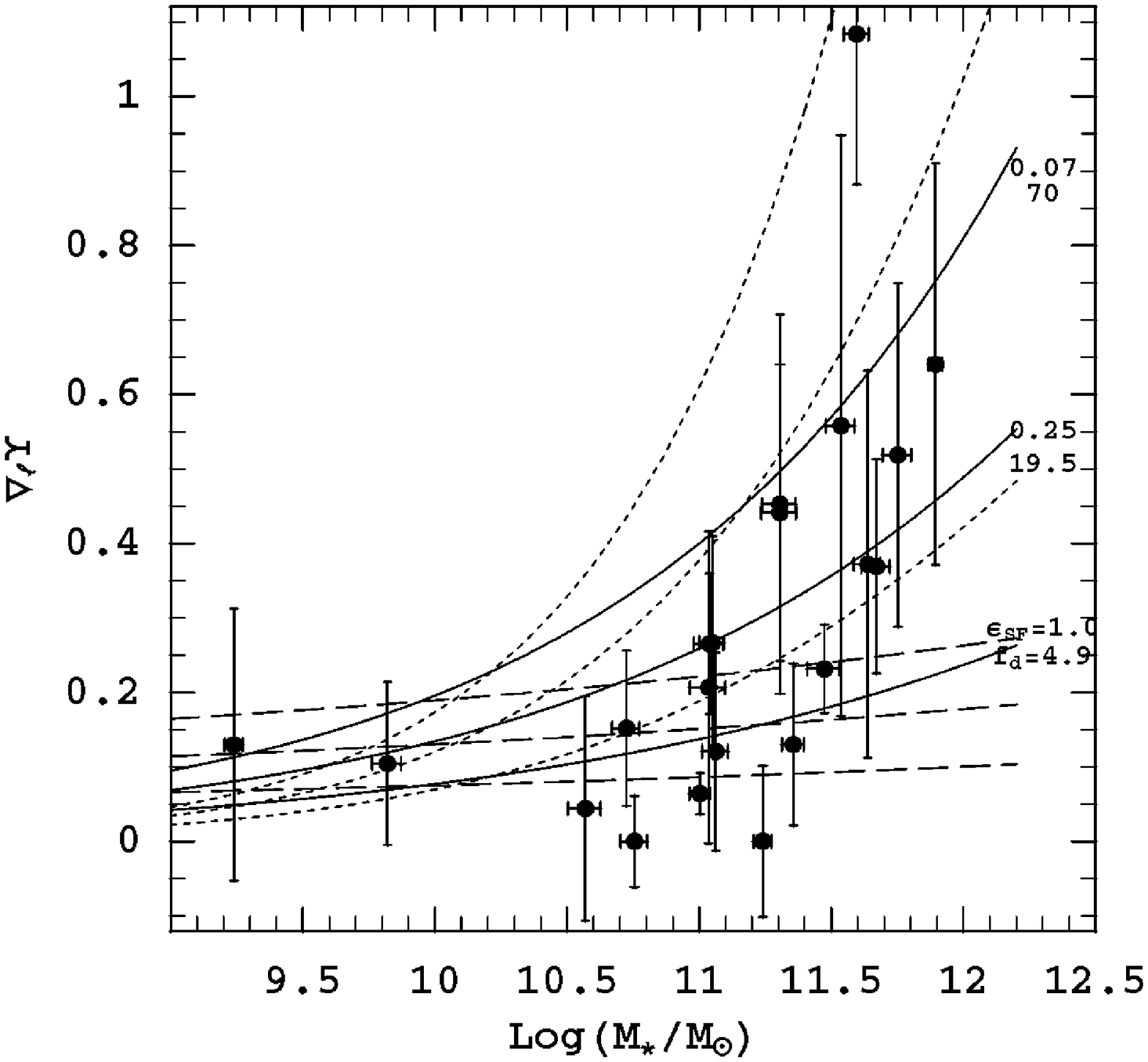,width=5.5cm,height=5.5cm,angle=-0}
%\hspace{-0.15cm} 
\epsfig{file=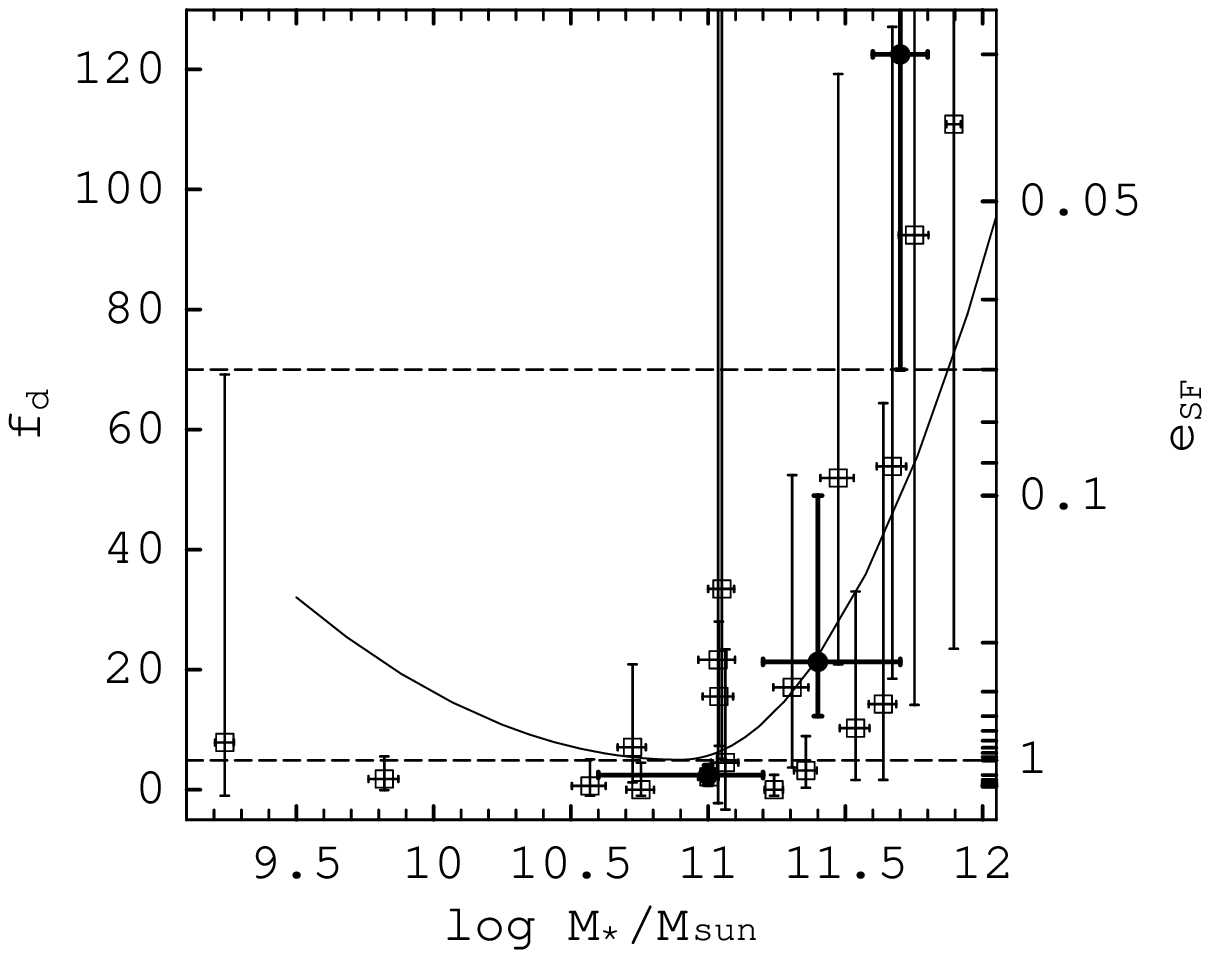,width=6.5cm,height=5cm,angle=-0}
 \caption{ Left: model \dML\ (lines) and observed \dML\ (dots) from N05. Right: inferred $f_{\rm d}=$ ($\epsilon_{\rm SF}$) for individual galaxies (squares with errorbars) as in N05 with overplotted smoothed $f_{\rm d}$ discussed in Section 3. See also text for details.}
       \label{fig3}
   \end{figure}

\begin{figure}[t]
   \centering
  \vspace{-0.4cm}
\epsfig{file=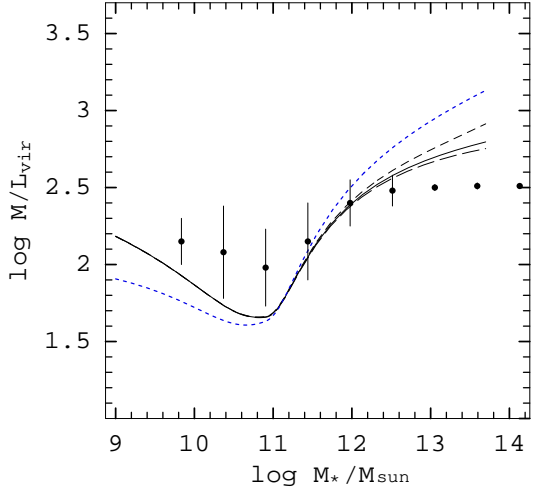,width=7cm,height=6.5cm,angle=-0}
 \caption{Virial M/L derived from the $f_{\rm d}$ smoothed function (solid line; long and short-dashed lines are uncertainties on the model) compared with the predictions from \cite{ben00} (data points with errorbars). Dotted blue line is the same as solid line but assuming $\Upsilon_*\propto M_*^{0.2}$ instead of $\Upsilon_*\propto M_*^{0.07}$ from Trujillo et al. (2004). See also text for details.}
       \label{fig4}
   \end{figure}

In Fig. \ref{fig3} (left) we compare the model \dML\ to the observed gradients of a sample of 21 early-type galaxies. 
%Solid lines are the prediction obtained assuming galaxies as luminous spheroids following an Hernquist (1990) profile embedded in NFW dark halos \footnote{\cite{nav04} provide a more detailed analysis of the density profiles, but using these profiles \dML\ by only $\sim$~10\%, which does not affect our conclusions.}. 
As discussed in N05, \dML\ are considered as a two parameter family models (stellar, $M_*$, and dark mass, $M_{\rm d}$ or $f_{\rm d}=$$M_{\rm d}/M_*$ as adopted in Fig. \ref{fig3}). The dark-to-luminous fraction, $f_{\rm d}$, can be interpreted in terms of star formation efficiency, $\epsilon_{\rm SF}=M_*/M_{\rm bar}$, i.e. the fraction of baryonic mass $M_{\rm bar}$ cooled in stars, assuming baryon conservation, such that  $\epsilon_{\rm SF}=4.9/f_{\mathrm d}$  (see N05 for further details).\\
In Fig. \ref{fig3}, the observed gradients do not seem to follow the simple increase of
\dML\ with $M_*$ that is expected for a ``universal'' $\epsilon_{\rm SF}$. This is made clearer in Fig. \ref{fig3} (right) where the inferred $\epsilon_{\rm SF}$ ($f_{\mathrm d}$) are plotted against the galaxy masses: massive (brightest) sample is broadly distributed within the physical meaningful $\epsilon_{\rm SF}$ range marked by the dashed lines, while less massive galaxies (included the R03 sample) have a very sharp distribution around a maximum $\epsilon_{\rm SF}$($>1$) values which is unphysical. The mass scale marking this ``dichotomy'' is $\log (M_0/M_{\odot})\sim 11.2$ [$M_0 \sim 1.6\times10^{11}
M_{\odot}$]. 
The overall trend seems to confirm other independent evidences that the star formation efficiency is a function of the halo masses with a minimum around $L_*$ galaxies (Benson et al. 2000, Dekel \& Bornboi 2004, see also Marinoni \& Hudson 2002, van den Bosch et al. 2003).
%: $\epsilon_{\rm SF}$ has a minimum around $L_*$ galaxies and decreases at low and high mass regimes because of the effect of feedbacks. 
The too high $\epsilon_{\rm SF}$ (too low $f_{\mathrm d}$) we have found around $\log (M_0/M_{\odot})\sim 10.5-11.0$ is still far to be understood and it remains a challenge for the theory.\\
Driven by the results of \cite{ben00} and \cite{dekxx}, we can interpolate a smoothed $f_{\mathrm d}$ function through the values plotted in Fig. \ref{fig3} of the form $f_{\mathrm d}=A (x-x_0)^2 + B$ where $x=\log (M_*/M_{\odot})$\footnote{The dependence of the $\epsilon_{\rm SF}$ on the mass can be obtained by the transformation $\epsilon_{\rm SF}=4.9/f_{\mathrm d}$ (see also N05).} 
. Forcing the fit to a more physical minimum (we take $f_{\mathrm d}=B=4.9$, i.e. $\epsilon_{\rm SF}=1$), the best fit (overplotted in Fig. \ref{fig3}, right panel) is found for $x_0=10.9$ and with $A=13.8$ for $x<11$ and $A=68$ for $x\ge 11$.
%had a double component with $A=$ and $B=$ for $\log (M_0/M_{\odot}) < 11.0$ and $A= ,~B=$ for $\log (M_0/M_{\odot})\ge 11.0$.\\ 
%\section{Implications}
%Here we want to check whether the $f_{\mathrm d}$ dependence on the mass discussed in the previous section is consistent with the expectation from semi-analytical models. 
Is this result consistent with semi-analytical N-body simulations?\\
Models from \cite{ben00} make explicit prediction of the M/L at the virial radius. 
%Up to date, these are the only suitable test of the scenario discussed above.
We use the toy-model illustrated in N05 (in their Sect. 3) in order to compute the M/L at the virial radius produced by the $f_{\mathrm d}$ function derived above. 
Indeed, by the definition of M/L:
\begin{equation} 
\frac{M}{L}=\Upsilon_* \frac{M_{\rm bar}+M_{\rm d}}{M_*}=\Upsilon_* \frac{M_{\rm bar}+M_{\rm d}}{\epsilon_{\rm SF}M_{\rm bar}}=5.9 \frac{\Upsilon_*}{\epsilon_{\rm SF}},
\label{MLdef} 
\end{equation}
where $M_*$ and $M_{\rm d}$ are referred to the virial radius, $\Upsilon_*$ is a constant stellar mass-to-light ratio (from population synthesis models) and we have used the cosmic baryon fraction $\Omega_{\mathrm{bar}}/\Omega_{\mathrm{tot}}=0.17$.
%(\cite{benn03}).\\
This shows that M/Ls at the virial radius depend on luminosity (and mass) through the quantities $\Upsilon_*$ and $\epsilon_{\rm SF}$.\\
There are indications that $\Upsilon_*$  has only a weak dependence on $M_*$ (Trujillo et al. 2004, we use their results in the following) which would leave $\epsilon_{\rm SF}$ as the main player of the virial $M/L$ dependence on the mass. Virial $M/L_{\rm B}$  predicted according with Eq. \ref{MLdef} are compared with results from \cite{ben00} in Fig. \ref{fig4}, where we have rescaled their halo masses to stellar masses trough our $f_{\rm d}$. There is a substantial agreement with semi-analytical model prediction except in the low mass regime where our $\epsilon_{\rm SF}$ estimates are based on poor data. But the overall behavior is reproduced. We also see that if we assume a stronger dependence of the $\Upsilon_*$ on the mass ($\Upsilon_*\propto M_*^{0.2}$), the trend of our predicted virial M/L is inconsistent with \cite{ben00}. 
\section{Conclusions}
The main conclusions are: 
1) for the moderate luminosity sample of R03, strong radial anisotropy does not seem to be an issue; 2) we have shown that the global dark-matter fraction in ellipticals is a smoothed function of the stellar mass, consistently with the predictions of semi-analytical models; 
%presented some evidences that strong radial anisotropy as addressed by Dekel et al. 2005 for explaining declining dispersion profiles of early-type (post-merger) galaxies, is poorly consistent with R03 data (when accounting for kurtosis).\\
%The global dark-matter fraction is a function of the stellar mass: following N05, we derived a smooth relation which allowed to estimate virial M/L consistent with semi-analytical models.\\
3) the low-M/L gradients found by R03 and N05, for intermediate luminosity systems,  imply too low dark-matter fraction challenging the standard $\Lambda CDM$, a problem that can be alleviated assuming low concentrated halos (R03, N05) or a lower $\sigma_8$ (Napolitano 2004).


\begin{thebibliography}{} 
%\bibitem[Bahcall, Lubin \& Dorman(1995)]{bahcall} Bahcall N.A., Lubin L.M., Dorma V., 1995, ApJ, 447, L81
%\bibitem[Bender et al.(1989)]{ben89} Bender R., Surma P., Doebereiner S., Moellenhoff C., Madejsky R., 1989  A\&A, 217, 35
%\bibitem[Bennett et al. (2003)]{benn03} Bennett C.L. et al.\ 2003, ApJs, 148, 1 
\bibitem[Benson et al. (2000)]{ben00} Benson A.J., Cole S., Frenk C.S., Baugh C.M., Lacey C.G., 2000, MNRAS, 311, 793  
%\bibitem[Bernardi et al.(2003)]{bernardi03} Bernardi M. et al., 2003, AJ, 125, 1849
%\bibitem[Bertin et al.(1994)]{bertin94} Bertin G. et al., 1994, A\&A, 292, 381
%\bibitem[Blumenthal et al.(1986)]{blum86} Blumenthal, G.~R., Faber, S.~M., Flores, R., \& Primack, J.R., 1986, ApJ, 301, 27 
%\bibitem[Borriello, Salucci \& Danese(2003)]{BSD03} Borriello A., Salucci P., Danese L., 2003, MNRAS, 341, 1109 
\bibitem[Bullock et al.(2001)]{bull01} Bullock J.S., et al., 2001, MNRAS, 321, 559 
\bibitem[Dekel \& Bornboi (2004)]{dekxx} Dekel, A., \& Birnboim, Y., 2005 [preprint:astro-ph/0412300]
\bibitem[Dekel et al. (2005)]{dekmam} Dekel, A., et al., 2005 [preprint:astro-ph/0501622]
%\bibitem[de Vaucouleurs(1948)]{dV48} de Vaucouleurs G., 1948, Ann. d'Astrophys., 11, 247
%\bibitem[Faber et al.(1997)]{fab97} Faber, S.M. et al., 1997, AJ, 114, 1771 
%\bibitem[Fabbiano(2003)]{gf03} Fabbiano, G. 2003, Adv. Space Res., 32, 2013
%\bibitem[Ford et al.(1996)]{16} Ford, H. C., et al. 1996, ApJ 472, 145 
%\bibitem[Gerhard et al.(2001)]{gerhard01} Gerhard O. et al., 2001, AJ, 121, 1936
%\bibitem[Gebhardt et al.(2000)]{17} Gebhardt K. et al., 2000, AJ, 119, 1157 
%\bibitem[Gebhardt et al.(2003)]{kg03} Gebhardt K. et al., 2003, ApJ, 583, 92 (G03)
%\bibitem[Gnedin et al.(2004)]{gne04} Gnedin O.Y., Kravtsov A.V., Klypin A.A., Nagai D., 2004, ApJ, submitted [preprint astro-ph/0406247)
%\bibitem[Graham et al.(2003)]{grahal03} Graham, A.~W., Erwin, P., Trujillo, I., \& Asensio Ramos, A., 2003, AJ, 125, 2951 
%\bibitem[Graham \& Guzm{\' a}n(2003)]{grahgu03} Graham, A.~W.~\&  Guzm{\' a}n, R., 2003, AJ, 125, 2936 
%\bibitem[Guzik \& Seljak(2002)]{guse02} Guzik J., Seljak U., 2002, MNRAS, 335, 311 
%\bibitem[Hernquist (1990)]{hern90} Hernquist L., 1990, ApJ, 356, 359 
%\bibitem[Kauffmann et al.(2003)]{kauffmann03} Kauffmann G. et al., 2003, MNRAS, 341, 33
\bibitem[Lokas \& Mamon (2003)]{loma03} Lokas, E.L. \& Mamon, G.A. 2001, MNRAS, 343, 401 
\bibitem[Marinoni \& Hudson (2002)]{mahu02} Marinoni, C.~\& Hudson, M.J., 2002, ApJ, 569, 101 
%\bibitem[Mathews \& Brighenti(2003)]{MBXr03} Mathews, W.~G.~\& Brighenti, F.\ 2003, \araa, 41, 191 
%\bibitem[Matsushita(2001)]{mats01} Matsushita, K. 2001, ApJ, 547, 693 
%\bibitem[M\'{e}ndez et al. (2001)]{mend} M\'{e}ndez, R.H. et al., 2001, ApJ, 563, 135 
\bibitem[Milgrom \& Sanders (2003)]{san03} Milgrom, M., \& Sanders, R.H., 2003, ApJL, 599, 25
%\bibitem[Napolitano et al.(2001)]{30} Napolitano, N.R., Arnaboldi, M., Freeman, K.C., \& Capaccioli, M. 2001, A\&A, 377, 784 
%\bibitem[Napolitano et al.(2002)]{nap02} Napolitano N.R., Arnaboldi M., Capaccioli M., 2002, A\&A, 383, 791 (NAC02)
\bibitem[Napolitano et al. (2005)]{nap05} Napolitano N.R., et al., 2005, MNRAS,
357, 691 (N05)
\bibitem[Napolitano (2004)]{nap04} Napolitano N.R., 2004. SISSA, Proceedings of Science, http://pos.sissa.it, p.67.1
\bibitem[Navarro et al. (1997)]{nfw97} Navarro J.F., Frenk C.S., White S.D., 1997, ApJ, 490, 493  (NFW)
%\bibitem[Navarro et al. (2004)]{nav04} Navarro J.F. et al., 2004, MNRAS, 349, 1039
%\bibitem[Peng, Ford \& Freeman(2004)]{peng04} Peng E.W., Ford H.C., Freeman K.C., 2004, ApJ, 602, 685 (P04)
\bibitem[Romanowsky et al. (2003)]{ral03} Romanowsky A.J. et al., 2003, Science, 301, 1696 (R03)
%\bibitem[Ryder et al.(2001)]{40} Ryder, S.D. et al. 2001, ApJ, 555, 232 
%\bibitem[Saglia et al.(1993)]{41} Saglia R.P. et al., 1993, ApJ, 403, 567 
%\bibitem[Shen et al.(2003)]{shen03} Shen S., Mo H.J., White S.D.M., Blanton M.R., Kauffmann G., Voges W., Brinkmann J., Csabai I., 2003, MNRAS, 343, 978 
\bibitem[Trujillo et al. 2004]{tbb04} Trujillo I., Burkert A.,  Bell E. F., 2004, ApJ, 600, L39
\bibitem[van den Bosch, Mo \& Yang (2003)]{vdB} van den Bosch F.C., Mo H.J., Yang X., 2003, MNRAS, 345, 923
%\bibitem[van den Bosch et al.(2004)]{vdBal04} van den Bosch F.C., Norberg P., Mo H.J., Yang X., 2004, MNRAS, 352, 1302
\bibitem[van der Marel \& Franx (1993)]{vdMF93} van der Marel, 
R.P., \& Franx, M.\ 1993, ApJ, 407, 525 
 \end{thebibliography}
\end{document}